\documentclass[runningheads]{llncs}
\usepackage[T1]{fontenc}
\usepackage[hidelinks]{hyperref}
\hypersetup{hypertexnames=false}
\usepackage{graphicx}
\usepackage{float}
\usepackage{placeins}
\usepackage{url}

\usepackage{cite}
\usepackage[table]{xcolor}
\usepackage{array}
\usepackage{booktabs}
\usepackage{multirow}
\usepackage{microtype}
\definecolor{FigSemanticRow}{HTML}{B5DDCB}
\definecolor{FigSpeakerRow}{HTML}{F6C7BA}
\definecolor{FigProsodyRow}{HTML}{FDD09E}
\definecolor{FigVoiceRow}{HTML}{B3D5F3}
\definecolor{FigAffectiveRow}{HTML}{FBD998}
\definecolor{FigSceneRow}{HTML}{C5C2E3}
\colorlet{SemanticTint}{FigSemanticRow!38!white}
\colorlet{SpeakerTint}{FigSpeakerRow!38!white}
\colorlet{ProsodyTint}{FigProsodyRow!38!white}
\colorlet{VoiceTint}{FigVoiceRow!38!white}
\colorlet{AffectiveTint}{FigAffectiveRow!38!white}
\colorlet{SceneTint}{FigSceneRow!38!white}
\newcommand{\schemacolorlegend}{%
\setlength{\fboxsep}{1.0pt}%
\resizebox{1.0\linewidth}{!}{\colorbox{SemanticTint}{Semantic Understanding} \colorbox{SpeakerTint}{Speaker Profile} \colorbox{ProsodyTint}{Prosody and Delivery} \colorbox{VoiceTint}{Voice Quality and Articulation} \colorbox{AffectiveTint}{Affective and Paralinguistic Cues} \colorbox{SceneTint}{Acoustic Scene}}}

\begin{document}

\title{SpeechAnnotator: A Context-Aware Multi-Agent Framework and Benchmark for Multidimensional Speech Annotation}
\titlerunning{SpeechAnnotator}

\author{Qirui Zhan\inst{1} \and
Shuiyuan Wang\inst{1} \and
Jingbin Hu\inst{1} \and
Haoyu Zhang\inst{1} \and
Xiaming Ren\inst{1} \and
Jinrui Liang\inst{1} \and
Chaoren Yu\inst{1} \and
Bengu Wu\inst{3} \and
Yunxiang Chen\inst{4} \and
Houdun Liu\inst{4} \and
Su Feng\inst{4} \and
Liumeng Xue\inst{2}\thanks{Corresponding authors.} \and
Lei Xie\inst{1}\textsuperscript{$\star$}}

\authorrunning{Q. Zhan et al.}

\institute{Northwestern Polytechnical University, China
\and
Nanjing University, China
\and
yutuzhineng, China
\and
Shenzhen Pimei Technology Co., Ltd., Guangdong, China\\
\email{zhanqirui@mail.nwpu.edu.cn, lmxue@nju.edu.cn, lxie@nwpu.edu.cn}}

\maketitle
\setcounter{footnote}{0}

\begin{abstract}
Recent controllable speech generation requires training data with fine-grained annotations of speaker traits, prosody, emotion, paralinguistic cues, acoustic scenes, and context. Existing workflows often rely on manual correction, paid hosted multimodal services, or fixed processing chains, which limits large-scale data processing through annotation cost, external-service dependence, or weak cross-stage recovery. We introduce \textbf{SpeechAnnotator}, a locally deployable, context-aware multi-agent framework built entirely from \textbf{open-source} models and tools. Supporting frontend modules first obtain speaker-aware segments and final segment transcripts, while prior evidence extractors attach heterogeneous segment-level cues. Three specialist agents then collaborate through shared state: the Planning Agent converts local audio evidence, speaker history, neighboring segments, and recording-level context into field-specific contracts; the Labeling Agent performs contract-guided multimodal prediction for directly observable attributes; and the Review Agent runs a bounded review loop that checks evidence support and cross-segment consistency, triggering relabeling only for unsupported or inconsistent fields. Existing evaluation resources are fragmented across isolated tasks, while most test sets cover only a narrow range of domains. To address these limitations, we introduce \textbf{SpeechAnnotator-Bench} (SA-Bench), containing 8.87 hours of human-annotated audio across nine source formats. We further propose \textbf{SpeechAnnotator-Eval} (SA-Eval), which separates Timeline-Eval for speaker-aware timeline recovery, Closed-Eval for finite-set attributes, and Open-Eval for open-ended attributes. Experiments and ablations show that SpeechAnnotator offers a locally deployable alternative to commercial audio-capable systems, with the bounded review loop improving multidimensional annotation through evidence- and context-aware field-level recovery. SA-Bench and the related code will be open-sourced\footnote{\url{https://github.com/ASLP-lab/SpeechAnnotator}}.
\end{abstract}

\section{Introduction}
Controllable speech generation has expanded the annotation requirements of speech data beyond transcripts and coarse event tags. Segment-level supervision must describe who is speaking, how an utterance is delivered, which acoustic scene surrounds it, and what the segment implies in context. These dimensions connect linguistic content, vocal realization, speaker traits, paralinguistic cues, and scene-level evidence.

Existing data construction pipelines address parts of this need but remain limited in scale, cost, or context. Human-centered workflows such as AudioSet~\cite{audioset} and NVSpeech~\cite{nvspeech} provide reliable labels but become costly when expanded to large-scale data annotation with richer schemas. Model-assisted resources such as WavCaps~\cite{wavcaps}, LTU-AS~\cite{ltu-as}, and VoiceSculptor~\cite{voicesculptor} reduce manual effort through large-language-model (LLM)-based rewriting, automatic speech recognition (ASR), alignment, acoustic feature extraction, proprietary multimodal analysis, or human calibration. VoiceSculptor further relies on low-level voice attributes from dedicated acoustic models rather than directly labeling the natural-language attribute schema studied here. Large-scale data processing with hosted application programming interfaces (APIs) may still incur recurring per-call cost and throughput constraints. MOSS-Audio~\cite{yang2026mossaudiotechnicalreport} offers content-adaptive segmentation and caption fusion, yet its annotation remains centered on individual segmented events. Long heterogeneous recordings require field-level evidence routing across local audio, speaker history, neighboring turns, and recording-level context.

Speech and audio benchmarks remain fragmented across task types. Speaker-diarization and audio-visual diarization datasets~\cite{aishell4,alimeeting,ami,ava-avd,callhome,dihard3,ego4d,msdwild,ramc,repere,voxconverse} provide timeline or person-presence testbeds in scenario-specific settings. MSU-Bench targets speaker-centric multi-speaker conversation~\cite{msu-bench}, while neighboring audio-language benchmarks cover instruction following or reasoning across speech, sound, and music~\cite{air-bench,mmau,mmar}. Multidimensional segment-level attributes, including speaker traits, prosody, paralinguistic cues, acoustic scene, and contextual inference, remain scattered rather than concentrated in one dataset. Evaluation must also separate lexical, speaker-attribution, and temporal-alignment errors from closed-set attribute matching and open-ended semantic comparison. Open-ended semantic comparison cannot be handled reliably by exact strings because valid annotations may differ in wording or granularity; representation-based semantic metrics such as BERTScore~\cite{bertscore} and Sentence-BERT~\cite{sentence-bert}, together with recent hidden-state-based ASR evaluation~\cite{llm-asr-eval}, motivate a deterministic hidden-state comparison space that captures paraphrastic similarity without introducing additional judge-model calls.

Taken together, the preceding challenges concern both costs and long-recording consistency: scalable multidimensional labeling must integrate heterogeneous evidence without relying on costly hosted services, while preserving coherent decisions across speakers, neighboring segments, and recording-level context. We address these challenges with \textbf{SpeechAnnotator}, a context-aware multi-agent annotation framework. The Planning Agent prepares compact field contracts, and the Labeling Agent produces initial annotations, while a bounded review loop provides the main recovery mechanism. The Review Agent checks consistency with retained evidence, speaker history, neighboring segments, and related fields, then writes targeted relabeling requests back to shared state. Only affected fields are revised, avoiding repeated timeline construction, prior extraction, and annotation of unrelated attributes. Supporting frontend modules construct the speaker-aware timeline and segment transcripts, prior extractors supply acoustic, affective, and sound-event evidence, and deterministic output processing normalizes and serializes reviewed annotations. We also introduce SpeechAnnotator-Bench and SpeechAnnotator-Eval for unified evaluation of timeline recovery, finite-set attributes, and open-ended semantic descriptions under one schema.

The main contributions are as follows:
(1) We propose SpeechAnnotator, an open-source context-aware multi-agent framework centered on a bounded review loop for evidence- and context-aware field-level revision. Lightweight field contracts route evidence for labeling, while supporting modules provide timeline construction, transcript refinement, prior evidence, and deterministic output normalization. An online demo is available at the SpeechAnnotator demo page\footnote{\url{https://zhanqirui.github.io/SpeechAnnotator-demo-page/}}.
(2) We introduce SpeechAnnotator-Bench (SA-Bench) and SpeechAnnotator-Eval (SA-Eval), which jointly assess speaker-aware timeline recovery, finite-set attribute prediction, and open-ended descriptive annotation under a unified schema. Open-Eval provides representation-based semantic scoring for natural-language attributes while keeping timeline, closed-set, and open-ended metrics separate.

\iffalse
\fi
\section{Context-Aware Multi-Agent Annotation Framework}

This section describes the design of SpeechAnnotator. It first defines the shared segment-level state that connects intermediate artifacts, then introduces the roles of the Planning, Labeling, and Review Agents, and finally details how supporting modules, prior evidence extractors, and the bounded review loop produce schema-aligned annotations.

\subsection{Overview and Shared State}
\begin{figure}[t]
\centering
\includegraphics[width=\textwidth]{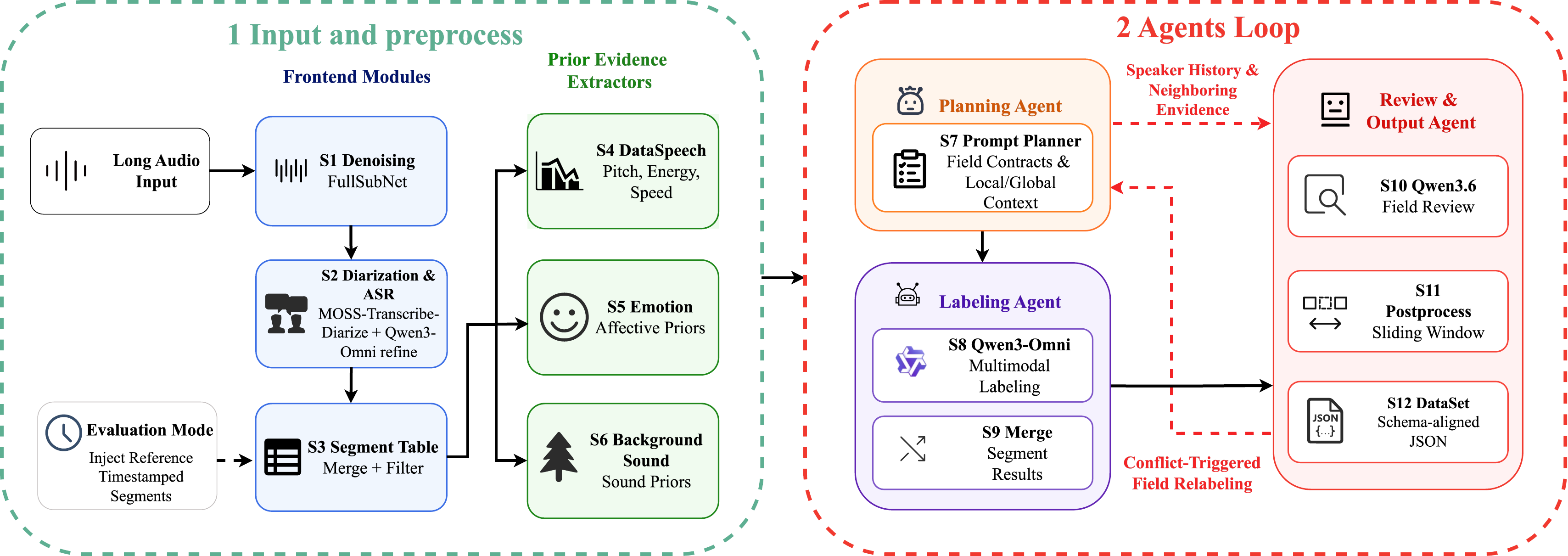}
    \caption{Overview of SpeechAnnotator. Supporting frontend modules construct segment records with final transcripts, while prior evidence extractors add field-relevant cues to shared state. The Planning Agent converts local and recording-level context into field-specific contracts for the Labeling Agent. The Review Agent verifies evidence and cross-segment consistency and triggers bounded relabeling of conflicting fields without restarting the supporting pipeline; deterministic output modules normalize and serialize reviewed annotations.
    }
\label{fig:pipeline-architecture}
\end{figure}

SpeechAnnotator formulates multidimensional audio-speech annotation as a controlled, segment-aligned framework for long-form audio. Artifacts---including enhanced waveform, speaker-aware timeline, segment audio, final transcripts, prior evidence, field contracts, predictions, reviewed records, and final JavaScript Object Notation (JSON) output---share a segment key. This shared state connects four workflow parts: frontend modules build timeline and segment records, prior extractors attach acoustic and semantic cues, the Planning, Labeling, and Review Agents coordinate field-level annotation and revision, and deterministic output modules normalize reviewed records into final JSON. The twelve-stage implementation combines frontend processing, prior extraction, multi-agent annotation, and deterministic output processing, as illustrated in Figure~\ref{fig:pipeline-architecture}.

\subsection{Agent Roles and Coordination Protocol}
The Planning Agent observes the segment table, transcript, prior evidence, global recording summary, speaker history, and neighboring segments. It decides which evidence is admissible for each field, how that evidence should be prioritized, which context scope is appropriate, and how potential conflicts should be presented, then writes a field-specific contract into shared state.

The Labeling Agent observes the original segment audio, transcript, selected evidence, permitted context, and field contract, and writes initial predictions for directly observed attributes into shared state. Caption and contextual inference are constructed later from the reviewed state. The Review Agent observes predictions, contracts, retained prior evidence, speaker history, neighboring-segment evidence, and related fields. It accepts sufficiently supported predictions or issues a field-specific revision request for targeted relabeling; review status and any revised prediction are then written back to shared state.

\subsection{Modules, Agents, and the bounded review loop}
Frontend modules create the segment backbone. FullSubNet~\cite{fullsubnet} enhances speech for robust segmentation and acoustic analysis. MOSS-Transcribe-Diarize~\cite{moss-transcribe-diarize} estimates speaker-homogeneous intervals, speaker identifiers, and initial transcripts from the enhanced waveform; Qwen3-Omni~\cite{qwen-omni} then refines the MOSS segment transcripts before they enter the segment table. A segment-table module reconciles boundaries, speakers, segment audio, and refined transcripts into canonical records; in the fixed-timeline attribute-evaluation setting, reference boundaries, reference speakers, and reference transcripts rebuild this table, while later attribute scoring remains independent of the predicted timeline. Prior evidence extractors then attach field-relevant evidence by segment key: DataSpeech~\cite{dataspeech} estimates acoustic statistics such as pitch, energy, loudness, and speed on enhanced segments, while Emotion2vec~\cite{emotion2vec} and sound event detection (SED)~\cite{pretrainedsed} operate on original audio to preserve affective, non-speech, and background-scene cues.

The Planning Agent observes the segment table, transcript, prior evidence, recording summary, speaker history, and neighboring segments. It writes a lightweight field-specific contract that records the annotation objective, admissible evidence, context scope, and conflict-handling constraints. The contract mechanism is used to route field-relevant evidence: speaker-profile fields may use same-speaker history, prosodic and voice-quality fields prioritize local audio and DataSpeech evidence, emotion may use nearby turns as weak support, and scene fields remain grounded in acoustic or SED evidence.

The Labeling Agent, implemented with Qwen3-Omni~\cite{qwen-omni}, receives the original segment audio, transcript, selected evidence, permitted context, and contract, then writes initial predictions for directly observed attributes to the shared state. The bounded review loop is applied after labeling rather than to the whole pipeline. The Review Agent, implemented with Qwen3.6-27B~\cite{qwen3.6-27b} and inspired by AudioToolAgent~\cite{audiotoolagent} and Audio-Mind~\cite{audio-mind}, verifies whether each prediction is supported by retained evidence and compatible with speaker history, neighboring segments, and related fields. Unsupported background sounds, emotion--prosody mismatches, speaker-trait drift, or contextually inconsistent supporting attributes are converted into field-specific revision requests. Only affected fields are relabeled with the original segment audio, transcript, selected evidence, and revision request; frontend processing, unrelated prior extraction, and accepted labels are kept fixed. The loop stops when reviewed fields are accepted or when the bounded revision budget is exhausted.

The bounded design matters because many annotation errors are neither syntactic failures nor global pipeline failures. A background-sound prediction may be copied from lexical content without acoustic support, an emotion label may conflict with prosody, or a speaker-profile attribute may drift when adjacent turns are processed independently. The loop expresses these cases as local revision actions rather than full regeneration. It also preserves an audit trail in shared state: each reviewed record keeps the accepted prediction, supporting evidence, or field-specific reason for relabeling. This makes revision controllable and compatible with deterministic JSON output.

After the loop, deterministic output processing standardizes speaker identifiers, regularizes speaker-level attributes, canonicalizes schema fields, constructs caption and contextual inference by summarizing the transcript, reviewed attributes, and retained context, and serializes the final JSON. This final stage handles format validation and schema normalization, whereas the review loop handles semantic support and contextual consistency for the directly predicted attributes.

\section{SA-Bench and SA-Eval}

This section introduces the benchmark and evaluation protocol used to assess SpeechAnnotator. It first describes SA-Bench construction and source-format coverage, then defines the 15-attribute schema and the three evaluation components that separately measure timeline recovery, closed-set attributes, and open-ended natural-language attributes.

\subsection{SA-Bench Construction and Coverage}
The proposed framework requires a benchmark that jointly evaluates speaker-aware timeline recovery and segment-level multidimensional annotation, rather than treating transcription, sound-event detection, and generic audio captioning as isolated tasks. Existing resources typically cover only subsets of these capabilities or adopt incompatible annotation schemas, making them unsuitable for evaluating the complete framework under a shared schema.

\begin{figure}[t]
\centering
\includegraphics[width=0.61\textwidth]{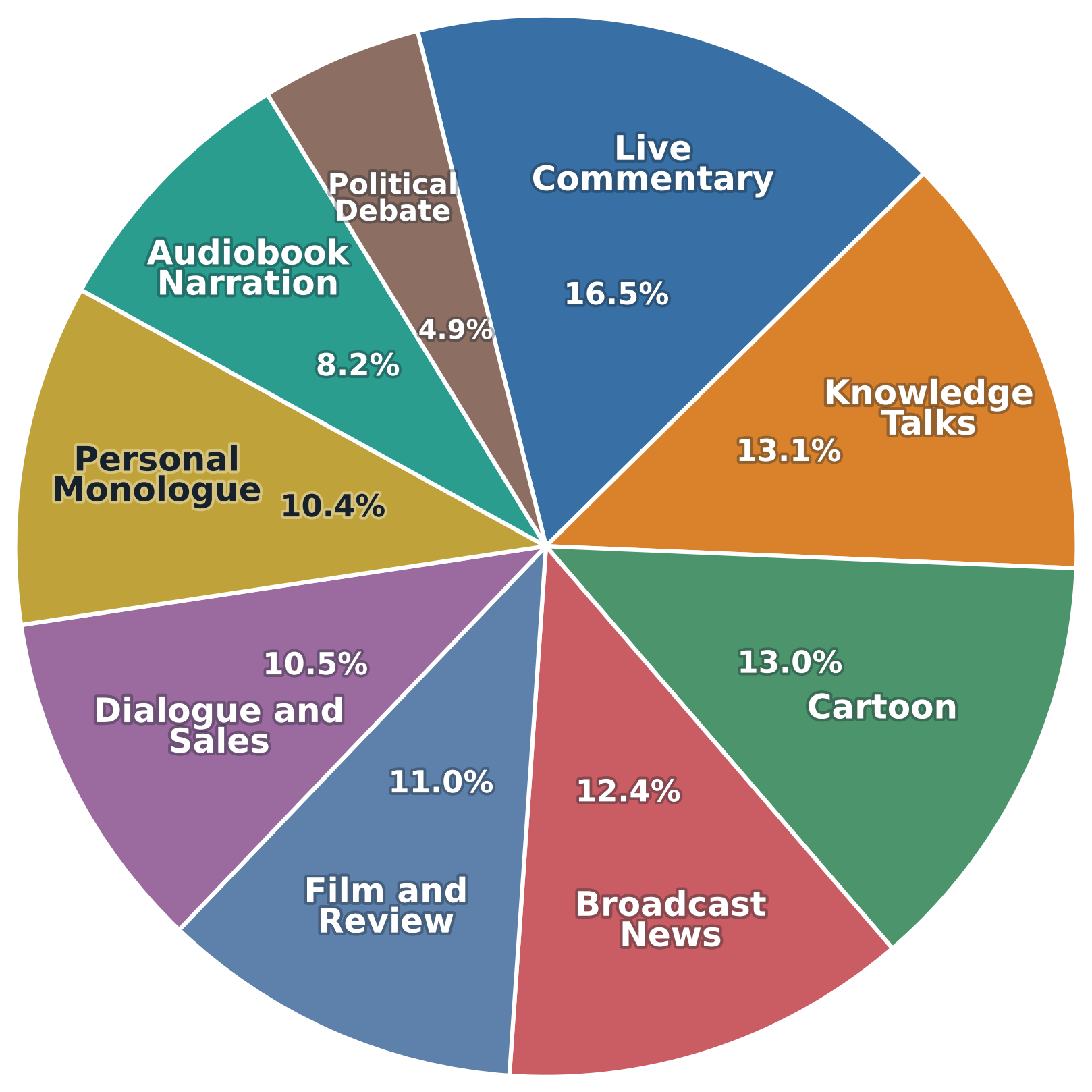}

\caption{Distribution of annotated duration across nine source formats in SA-Bench. Slice area indicates each format's share of the 8.87-hour reference timeline.}
\label{fig:SA-Bench-distribution}
\end{figure}
\FloatBarrier

We introduce SA-Bench, whose reference annotations span 8.87 hours of Chinese-dominant audio across nine source formats: Live Commentary, Knowledge Talks, Cartoon, Broadcast News, Film and Review, Dialogue and Sales, Personal Monologue, Audiobook Narration, and Political Debate. These formats are defined by how speech is produced and organized, rather than by a single topical taxonomy, because format diversity exposes different speaker-turn patterns, speaking styles, background conditions, and discourse structures. For each clip, the annotated duration is defined by its final reference timestamp, which also determines the common scoring range for timeline recovery and attribute prediction.

Annotation followed a five-annotator, two-inspector protocol. Five trained annotators independently labeled segment boundaries, speaker identifiers, transcripts, and multidimensional schema attributes. Two inspectors subsequently reviewed all annotations, returning inconsistent or substandard records for revision.

Figure~\ref{fig:SA-Bench-distribution} reports the distribution of the nine source formats. Live Commentary is the largest group, covering 1.46 hours (16.5\%), followed by Knowledge Talks with 1.16 hours (13.1\%), Cartoon with 1.16 hours (13.0\%), and Broadcast News with 1.10 hours (12.4\%). The remaining formats cover film and review material, dialogue and sales interactions, personal monologues, audiobook narration, and political debate, giving the benchmark broad source-format coverage without a single dominant group.

\begin{figure}[ht]
\centering
\includegraphics[width=0.9\textwidth]{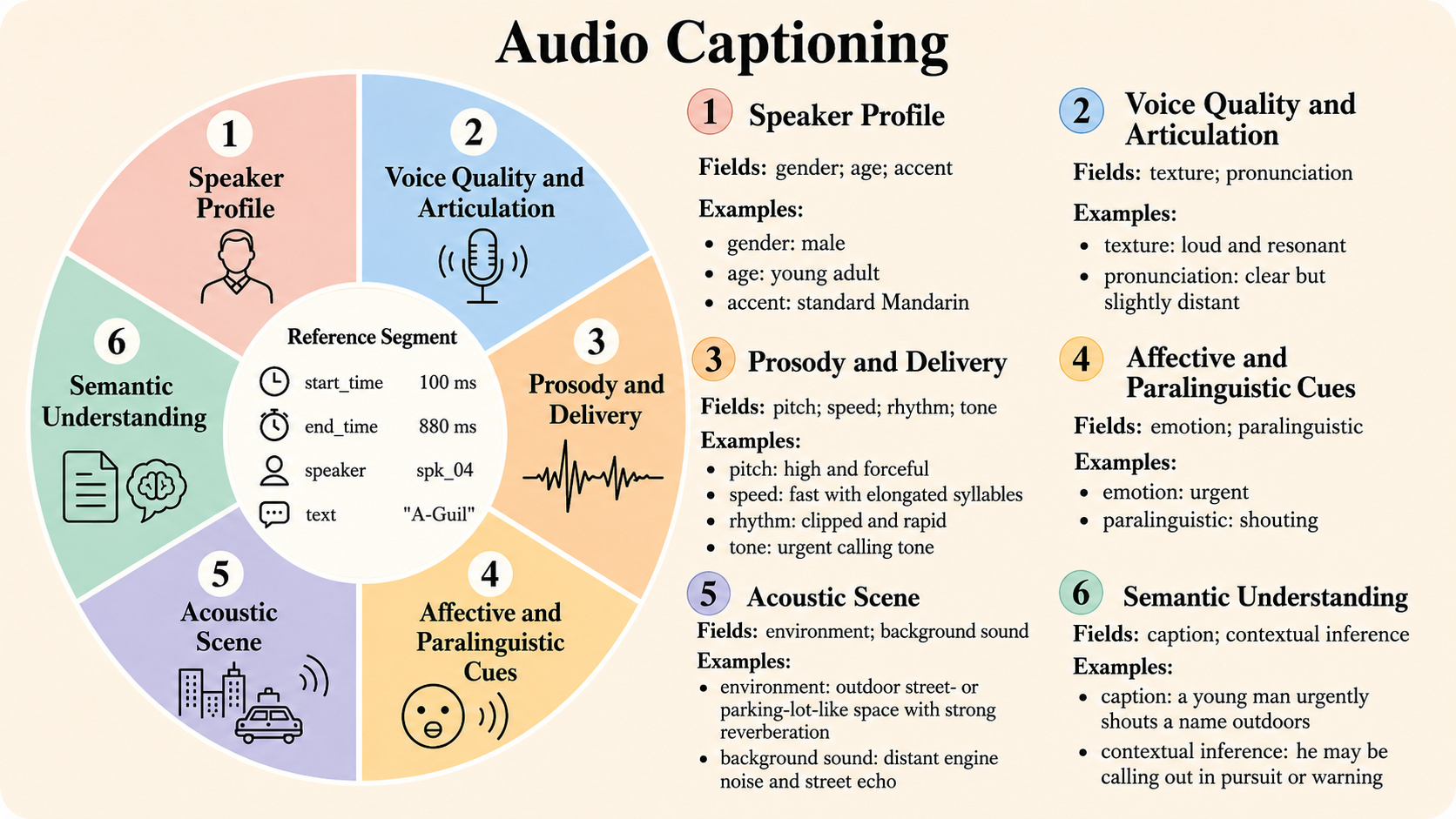}
\caption{Six evidence-oriented groups in the SpeechAnnotator annotation schema: Semantic Understanding, Speaker Profile, Prosody and Delivery, Voice Quality and Articulation, Affective and Paralinguistic Cues, and Acoustic Scene.}
\label{fig:schema-classification}
\end{figure}
\FloatBarrier

\subsection{Annotation Schema and Evaluation Protocol}
Based on evidence source and annotation role, the 15 evaluated attributes are first organized into six evidence-oriented groups, as illustrated in Figure~\ref{fig:schema-classification}. Semantic Understanding contains caption and contextual inference. Speaker Profile contains perceived gender, perceived age, and accent. Prosody and Delivery contains pitch, speed, rhythm, and tone. Voice Quality and Articulation contains texture and pronunciation. Affective and Paralinguistic Cues contains emotion and paralinguistic cues. Acoustic Scene contains environment and background sound. Each reference segment is annotated with temporal boundaries, a speaker identifier, a transcript, and values for all 15 attributes.

The 15 evaluated attributes form the attribute set $\mathcal{F}$. Segment boundaries, speaker identifiers, and transcripts are treated as structural fields and are not included in $\mathcal{F}$. For attribute evaluation, perceived gender and perceived age form the closed set $\mathcal{C}$, while the remaining attributes are natural-language descriptions and form the open-ended set $\mathcal{O}=\mathcal{F}\setminus\mathcal{C}$.

SA-Eval comprises three complementary components. On the Chinese-dominant benchmark, Timeline-Eval evaluates lexical content, speaker attribution, and temporal alignment using character error rate (CER), concatenated minimum-permutation CER (\textit{cp}CER), and time-constrained minimum-permutation CER (\textit{tcp}CER). Closed-Eval evaluates finite-set attributes under the fixed-timeline setting, whereas Open-Eval evaluates open-ended attributes using semantic similarity between Qwen3.6-27B representations. These components are reported separately rather than combined into a single end-to-end score.

The predicted speaker-aware timeline is evaluated directly using Timeline-Eval. Attribute evaluation follows a fixed-timeline setting: each reference segment provides its temporal boundaries, speaker identifier, and transcript, while the system predicts the attributes in $\mathcal{F}$. Missing or unparsable predictions remain in the evaluation denominator and receive a score of zero.

\subsection{Timeline-Eval: Transcription and Speaker-Aware Timing}
Because SA-Bench is Chinese-dominant, Timeline-Eval uses character-level error rates. Standard character error rate (CER), $\mathrm{CER}=(S+D+I)/N$, pools all characters and ignores speaker labels and temporal information.

Concatenated minimum-permutation CER (\textit{cp}CER) additionally accounts for speaker attribution. Let $R$ and $H$ denote the sets of reference and hypothesized speakers, respectively. After concatenating utterances from each speaker, the evaluator searches over admissible speaker mappings $\pi \in \Pi(H,R)$:

\begin{equation}
\label{eq:cpcer}
\textit{cp}\mathrm{CER}
=
\min_{\pi \in \Pi(H,R)}
\frac{
\sum_{r \in R}
\left(S^{\mathrm{cat}}_{r,\pi}+D^{\mathrm{cat}}_{r,\pi}+I^{\mathrm{cat}}_{r,\pi}\right)
}{
\sum_{r \in R} N_r
}.
\end{equation}

Time-constrained minimum-permutation CER (\textit{tcp}CER) further constrains the edit operations using character timestamps, thereby requiring both speaker-consistent and temporally valid alignments. Together, CER, \textit{cp}CER, and \textit{tcp}CER distinguish lexical errors from speaker-attribution and temporal-alignment errors.

\subsection{Closed-Eval and Open-Eval}

For attribute evaluation, the system receives the reference segment sequence
$G=\{g_i\}_{i=1}^{N}$. For segment $i$ and attribute $f \in \mathcal{F}$, let $y_{i,f}$ and $\hat{y}_{i,f}$ denote the reference and predicted values, respectively. A field-specific canonicalizer $c_f(\cdot)$ normalizes punctuation, case, spacing, aliases, and empty-value markers.

For each closed-set attribute $f \in \mathcal{C}$, Closed-Eval uses canonicalized exact matching as its evaluation metric:
\begin{equation}
s_{i,f}^{\mathrm{closed}}
=
\mathbf{1}
\left[
c_f(\hat{y}_{i,f})
=
c_f(y_{i,f})
\right].
\end{equation}

For open-ended attributes, exact string matching is overly restrictive because semantically equivalent annotations may differ in wording, granularity, or syntactic form. Open-Eval therefore formulates evaluation as reference-based semantic matching in a shared representation space.

Transformer hidden states provide contextualized representations in which each token is encoded jointly with its surrounding linguistic context. Prior representation-based methods have shown that such representations can capture semantic correspondence beyond surface-form overlap~\cite{bertscore,sentence-bert}. Recent work has further demonstrated that hidden states extracted from decoder-only generative LLMs can serve as effective semantic representations for evaluating ASR outputs under simple pooling and similarity functions~\cite{llm-asr-eval}. Motivated by these findings, Open-Eval uses the hidden states of Qwen3.6-27B to encode both reference and predicted attribute values.

Given an input text $x$, its last-layer hidden states $\mathbf{h}^{\mathrm{last}}_t$ are mean-pooled using the attention mask $m_t$ and then $\ell_2$-normalized:
\begin{equation}
\mathbf{e}(x)
=
\mathrm{Normalize}
\left(
\frac{\sum_{t=1}^{T} m_t \mathbf{h}^{\mathrm{last}}_t}
{\sum_{t=1}^{T} m_t}
\right).
\end{equation}

Open-Eval uses clipped cosine similarity as the evaluation metric for open-ended attributes:
\begin{equation}
s_{i,f}^{\mathrm{open}}
=
\max
\left(
0,\,
\cos
\left(
\mathbf{e}(c_f(\hat{y}_{i,f})),
\mathbf{e}(c_f(y_{i,f}))
\right)
\right),
\qquad
f \in \mathcal{O}.
\end{equation}

Mean pooling aggregates contextualized token states into a representation of the complete attribute description, while cosine similarity measures semantic proximity independently of representation magnitude. This formulation allows predictions with different surface realizations but similar semantic content to receive comparable scores. Open-Eval does not invoke Qwen3.6-27B as a generative judge or use review prompts; it uses only frozen hidden-state representations. The same canonicalization rules, pooling operation, and similarity function are applied to SpeechAnnotator and all baselines, ensuring a consistent field-wise comparison.

\section{Experiments}

\subsection{Evaluation Setup}
All systems are evaluated on SA-Bench as a held-out test set without any training or adaptation. They receive the same recordings and target output schema.

We compare SpeechAnnotator with four commercial, audio-capable multimodal systems. Gemini 2.5 Pro~\cite{gemini25pro} and Gemini 3.1 Pro Preview\footnote{\url{https://ai.google.dev/gemini-api/docs/models/gemini-3.1-pro-preview}} accept audio input through the Gemini API; Qwen3.5-Omni-plus~\cite{qwen35omni} is provided through Alibaba Cloud Model Studio; and Seed2.0 Lite~\cite{seed20lite} is ByteDance Seed's Lite multimodal model. These comparison systems are evaluated as black boxes using only audio inputs, prompts, controlled segment information, and returned text or JSON outputs.

We evaluate all systems under two complementary settings. The timeline-recovery setting assesses diarization-aware transcription directly from the original recordings. In Table~\ref{tab:timeline-results}, the SpeechAnnotator row uses MOSS-predicted boundaries and speaker labels together with Qwen3-Omni-refined MOSS transcripts, whereas the MOSS-Transcribe-Diarize row reports the standalone MOSS diarized transcript output. Each commercial system receives the same recording and a prompt requesting a diarized transcript in the target schema. Because SA-Bench is Chinese-dominant, Table~\ref{tab:timeline-results} reports character-level CER, \textit{cp}CER, and \textit{tcp}CER for transcription, speaker-attribution, and temporal-alignment performance.

The fixed-timeline setting evaluates multidimensional attribute annotation independently of timeline-recovery errors. All systems receive the same reference segment boundaries, speaker identifiers, and reference transcripts. In each run, each commercial baseline is prompted once per recording to predict all 15 attributes for these segments. SpeechAnnotator builds the segment table from the reference timeline and reference transcripts, replacing its predicted-transcript path in this controlled setting, and then runs speech enhancement, prior extraction, context-aware planning, multimodal labeling, review, and postprocessing. The returned records are parsed into the target annotation schema and compared with the reference segment by segment: perceived gender and perceived age are scored by Closed-Eval after canonicalization, and the remaining attributes are scored by Open-Eval. Each system is run five times under the same fixed-timeline inputs. Table~\ref{tab:field-results} therefore isolates attribute annotation from transcription and diarization quality.

\subsection{Main Results}
Table~\ref{tab:timeline-results} compares speaker-aware timeline recovery with character-level metrics. SpeechAnnotator obtains the lowest CER and \textit{tcp}CER, while Qwen3.5-Omni-plus~\cite{qwen35omni} obtains the lowest \textit{cp}CER. Compared with standalone MOSS-Transcribe-Diarize, SpeechAnnotator improves CER from 12.84\% to 12.42\% and \textit{tcp}CER from 34.08\% to 33.69\% after refining MOSS segment transcripts with Qwen3-Omni. Its \textit{tcp}CER is 19.92 points below the strongest commercial baseline, Seed2.0 Lite~\cite{seed20lite}.

\FloatBarrier
\begin{table}[ht!]

\caption{Speaker-aware timeline results on SA-Bench (\%). Lower is better; each entry is mean $\pm$ standard deviation over five runs. Metrics are character-level CER, \textit{cp}CER, and \textit{tcp}CER. Non-SpeechAnnotator rows are ordered by \textit{tcp}CER, with SpeechAnnotator placed last.}
\label{tab:timeline-results}
\centering
\small
\begin{tabular}{l@{\hspace{2.4em}}c@{\hspace{2.4em}}c@{\hspace{2.4em}}c}
\hline
System & CER$\downarrow$ & \textit{cp}CER$\downarrow$ & \textit{tcp}CER$\downarrow$ \\
\hline
Gemini 2.5 Pro & 20.51$\pm$0.71 & 37.80$\pm$1.05 & 69.59$\pm$2.31 \\
Gemini 3.1 Pro Preview & 23.06$\pm$0.82 & 42.58$\pm$1.22 & 67.18$\pm$2.08 \\
Qwen3.5-Omni-plus & 14.17$\pm$0.46 & \textbf{26.31$\pm$0.73} & 56.25$\pm$1.84 \\
Seed2.0 Lite & 13.90$\pm$0.58 & 33.62$\pm$0.96 & 53.61$\pm$1.67 \\
MOSS-Transcribe-Diarize & \underline{12.84$\pm$0.18} & 28.70$\pm$0.31 & \underline{34.08$\pm$0.42} \\
SpeechAnnotator & \textbf{12.42$\pm$0.18} & \underline{28.16$\pm$0.18} & \textbf{33.69$\pm$0.12} \\
\hline
\end{tabular}
\end{table}

Table~\ref{tab:field-results} reports attribute-annotation performance under the fixed-timeline setting. At the segment level, each closed-set attribute is assigned a canonicalized exact-match score, whereas each open-ended attribute is assigned a clipped semantic-similarity score; both scores lie in $[0,1]$. For each attribute, the segment-level scores are first averaged over all reference segments, and each table entry reports the mean and standard deviation of the resulting attribute score across five runs, expressed as percentages. The field rows list the 15 attributes directly, with row colors indicating the six schema categories in Figure~\ref{fig:schema-classification}; Average is the unweighted mean of all 15 field means. Canonical null labels, including no-event and none-style values, are retained and scored after canonicalization. Missing, unparsable, or failed outputs remain in the denominator and receive a score of zero.

\begin{table}[t]
\caption{Attribute scores under the fixed-timeline and reference-transcript setting (\%). Each entry is mean $\pm$ standard deviation over five runs. Field rows are shaded using the six schema colors defined in Figure~\ref{fig:schema-classification}. Bold and underlined values mark the best and second-best systems on mean scores.}
\label{tab:field-results}
\centering
\small
\setlength{\tabcolsep}{1.8pt}
\renewcommand{\arraystretch}{0.96}
\makebox[\textwidth][c]{\resizebox{1.0\textwidth}{!}{%
\begin{tabular}{@{}l@{\hspace{0.35em}}ccccc@{}}
\toprule
Field & SpeechAnnotator & Gemini 2.5 Pro & Gemini 3.1 Pro Preview & Qwen3.5-Omni-plus & Seed2.0 Lite \\
\midrule
Average & \underline{81.27$\pm$0.34} & \textbf{81.73$\pm$1.18} & 78.25$\pm$1.29 & 72.69$\pm$1.58 & 79.40$\pm$1.11 \\
\midrule
\rowcolor{SemanticTint} Caption & 87.06$\pm$0.18 & \textbf{92.80$\pm$0.72} & \underline{90.30$\pm$0.83} & 89.80$\pm$0.91 & 85.20$\pm$0.78 \\
\rowcolor{SemanticTint} Contextual inference & 73.60$\pm$0.31 & \textbf{86.40$\pm$1.18} & 81.70$\pm$1.07 & 78.00$\pm$1.26 & \underline{83.10$\pm$0.94} \\
\midrule
\rowcolor{SpeakerTint} Perceived gender & \underline{93.43$\pm$0.28} & 90.20$\pm$1.03 & 91.20$\pm$1.09 & \textbf{95.10$\pm$0.80} & 85.40$\pm$1.42 \\
\rowcolor{SpeakerTint} Perceived age & 75.57$\pm$0.48 & \underline{86.10$\pm$1.64} & 77.20$\pm$1.71 & 65.70$\pm$2.04 & \textbf{92.20$\pm$1.18} \\
\rowcolor{SpeakerTint} Accent & 95.27$\pm$0.22 & \textbf{96.90$\pm$0.64} & 88.90$\pm$1.36 & 93.90$\pm$0.95 & \underline{95.70$\pm$0.71} \\
\midrule
\rowcolor{ProsodyTint} Pitch & \textbf{86.87$\pm$0.20} & \underline{85.70$\pm$0.82} & 76.60$\pm$1.35 & 68.90$\pm$1.58 & 75.90$\pm$1.16 \\
\rowcolor{ProsodyTint} Speed & \textbf{92.30$\pm$0.17} & 62.10$\pm$1.72 & \underline{70.70$\pm$1.45} & 50.20$\pm$2.08 & 66.30$\pm$1.63 \\
\rowcolor{ProsodyTint} Rhythm & \textbf{85.49$\pm$0.24} & 76.80$\pm$1.13 & 76.50$\pm$1.26 & 63.10$\pm$1.84 & \underline{77.40$\pm$1.05} \\
\rowcolor{ProsodyTint} Tone & \textbf{82.34$\pm$0.26} & \underline{80.70$\pm$1.04} & 77.30$\pm$1.28 & 69.00$\pm$1.66 & 76.60$\pm$1.21 \\
\midrule
\rowcolor{VoiceTint} Texture & \textbf{85.08$\pm$0.21} & 71.80$\pm$1.22 & \underline{75.70$\pm$1.09} & 62.10$\pm$1.73 & 73.90$\pm$1.14 \\
\rowcolor{VoiceTint} Pronunciation & \textbf{87.89$\pm$0.19} & 77.90$\pm$1.18 & 77.50$\pm$1.22 & 62.20$\pm$1.81 & \underline{80.40$\pm$0.96} \\
\midrule
\rowcolor{AffectiveTint} Emotion & 62.85$\pm$0.49 & \textbf{73.90$\pm$1.46} & 63.00$\pm$1.37 & \underline{66.00$\pm$1.51} & 64.60$\pm$1.30 \\
\rowcolor{AffectiveTint} Paralinguistic cues & \textbf{89.19$\pm$0.25} & \underline{89.10$\pm$0.91} & 83.00$\pm$1.22 & 81.00$\pm$1.34 & 87.70$\pm$0.98 \\
\midrule
\rowcolor{SceneTint} Environment & 55.04$\pm$0.68 & 72.90$\pm$1.54 & \textbf{74.00$\pm$1.33} & 71.20$\pm$1.71 & \underline{73.30$\pm$1.25} \\
\rowcolor{SceneTint} Background sound & 67.07$\pm$0.71 & \textbf{82.60$\pm$1.37} & 70.10$\pm$1.59 & \underline{74.10$\pm$1.66} & 73.30$\pm$1.28 \\
\bottomrule
\end{tabular}
}}
\vspace{-0.8mm}
{\tiny\parbox{1.0\textwidth}{\centering\schemacolorlegend}}
\end{table}
\FloatBarrier

SpeechAnnotator achieves an attribute macro average of 81.27\%, ranking second overall and remaining 0.46 points below Gemini 2.5 Pro~\cite{gemini25pro}. It obtains the highest score on seven of the 15 fields: tone, pitch, speed, rhythm, texture, pronunciation, and paralinguistic cues. At the category level, SpeechAnnotator ranks first on Prosody and Delivery and Voice Quality and Articulation, with category means of 86.75\% and 86.48\%, respectively. Its largest field-level gains over the best comparison result in each field occur on speed, texture, rhythm, and pronunciation, with margins of 21.60, 9.38, 8.09, and 7.49 points. These gains are concentrated in prosody, delivery, voice quality, and articulation.

The grouped results also reveal limitations. Acoustic Scene is weakest because environment scores 55.04\%, trailing Gemini 3.1 Pro Preview by 18.96 points; short speech-centered segments often contain weak scene cues, so environment inference can drift toward transcript semantics. Emotion remains low because subtle or mixed affect is not always clear from local prosody, and perceived age is 16.63 points below Seed2.0 Lite~\cite{seed20lite}. Thus, the cross-system comparison demonstrates overall competitiveness but does not isolate the contributions of context-aware planning, prior evidence, or the review loop, which are examined through controlled ablations.

Overall, SpeechAnnotator remains competitive with the commercial comparison systems, pairing strong timeline recovery with the best performance on seven of the 15 evaluated fields and the second-highest attribute macro average.
\subsection{Ablation}
Table~\ref{tab:ablation-results} reports three controlled ablations under the same fixed-timeline setting as Table~\ref{tab:field-results}. The \emph{w/o Loop} variant removes the Review Agent and targeted field-level relabeling, while retaining deterministic validation and postprocessing. The \emph{w/o Context} variant restricts the Planning Agent's observation by removing the global recording summary, speaker history, and neighboring-segment evidence from context-aware planning, while retaining segment-level prior-extractor outputs, original audio inputs, multimodal labeling, and postprocessing. The \emph{w/o Priors} variant removes heterogeneous evidence from DataSpeech, Emotion2vec, and the SED module, while retaining the segment table, transcript, original audio, contextual observation, multimodal labeling, review, and postprocessing. These ablations intervene on three complementary components of agent collaboration: the contextual observation available to the Planning Agent, the external evidence available to planning and labeling, and the revision action provided by the Review Agent.

\begin{table}[t]
\caption{Ablation results under the fixed-timeline setting (\%). Each entry is mean $\pm$ standard deviation over five runs. Field rows are shaded using the six schema colors defined in Figure~\ref{fig:schema-classification}. Bold and underlined values mark the best and second-best variants on mean scores.}
\label{tab:ablation-results}
\centering
\small
\setlength{\tabcolsep}{2.4pt}
\renewcommand{\arraystretch}{0.94}
\makebox[\textwidth][c]{\resizebox{0.98\textwidth}{!}{%
\begin{tabular}{@{}l@{\hspace{0.35em}}cccc@{}}
\toprule
Field & SpeechAnnotator & w/o Loop & w/o Context & w/o Priors \\
\midrule
Average & \textbf{81.27$\pm$0.34} & \underline{80.42$\pm$0.39} & 79.79$\pm$0.48 & 80.07$\pm$0.42 \\
\midrule
\rowcolor{SemanticTint} Caption & \underline{87.06$\pm$0.18} & 85.27$\pm$0.21 & 82.29$\pm$0.34 & \textbf{87.16$\pm$0.22} \\
\rowcolor{SemanticTint} Contextual inference & \textbf{73.60$\pm$0.31} & \underline{73.54$\pm$0.35} & 69.77$\pm$0.48 & 73.51$\pm$0.36 \\
\midrule
\rowcolor{SpeakerTint} Perceived gender & \textbf{93.43$\pm$0.28} & 91.52$\pm$0.36 & \underline{92.52$\pm$0.41} & 92.39$\pm$0.39 \\
\rowcolor{SpeakerTint} Perceived age & \textbf{75.57$\pm$0.48} & 72.98$\pm$0.70 & 72.20$\pm$0.76 & \underline{74.87$\pm$0.66} \\
\rowcolor{SpeakerTint} Accent & \textbf{95.27$\pm$0.22} & \underline{94.38$\pm$0.26} & 92.33$\pm$0.45 & 93.65$\pm$0.31 \\
\midrule
\rowcolor{ProsodyTint} Pitch & \textbf{86.87$\pm$0.20} & \underline{85.59$\pm$0.24} & 84.97$\pm$0.36 & 84.80$\pm$0.25 \\
\rowcolor{ProsodyTint} Speed & \textbf{92.30$\pm$0.17} & 90.51$\pm$0.28 & \underline{92.00$\pm$0.31} & 89.10$\pm$0.23 \\
\rowcolor{ProsodyTint} Rhythm & \underline{85.49$\pm$0.24} & 84.47$\pm$0.27 & \textbf{85.62$\pm$0.33} & 85.47$\pm$0.28 \\
\rowcolor{ProsodyTint} Tone & \underline{82.34$\pm$0.26} & \textbf{82.40$\pm$0.30} & 81.38$\pm$0.39 & 81.59$\pm$0.34 \\
\midrule
\rowcolor{VoiceTint} Texture & \textbf{85.08$\pm$0.21} & \underline{84.97$\pm$0.24} & 84.67$\pm$0.32 & 84.24$\pm$0.30 \\
\rowcolor{VoiceTint} Pronunciation & \underline{87.89$\pm$0.19} & \textbf{88.08$\pm$0.23} & 87.49$\pm$0.30 & 87.62$\pm$0.27 \\
\midrule
\rowcolor{AffectiveTint} Emotion & \textbf{62.85$\pm$0.49} & 61.58$\pm$0.55 & \underline{62.58$\pm$0.61} & 60.41$\pm$0.58 \\
\rowcolor{AffectiveTint} Paralinguistic cues & \textbf{89.19$\pm$0.25} & \underline{88.80$\pm$0.29} & 87.07$\pm$0.43 & 88.20$\pm$0.35 \\
\midrule
\rowcolor{SceneTint} Environment & \underline{55.04$\pm$0.68} & 54.90$\pm$0.63 & \textbf{55.72$\pm$0.71} & 53.07$\pm$0.76 \\
\rowcolor{SceneTint} Background sound & \underline{67.07$\pm$0.71} & \textbf{67.30$\pm$0.78} & 66.27$\pm$0.86 & 64.97$\pm$0.82 \\
\bottomrule
\end{tabular}
}}
\vspace{-0.8mm}
{\tiny\parbox{0.98\textwidth}{\centering\schemacolorlegend}}
\vspace{-1.5mm}
\end{table}

Removing the review loop reduces the attribute macro average from 81.27\% to 80.42\%. The largest category-level decreases appear in Speaker Profile, Prosody and Delivery, Semantic Understanding, and Affective and Paralinguistic Cues, which drop by 1.80, 1.01, 0.92, and 0.83 points, respectively. At the field level, the most affected attributes are perceived age, perceived gender, caption, speed, pitch, and emotion, with decreases of 2.59, 1.91, 1.79, 1.79, 1.28, and 1.27 points. The macro-average standard deviation also increases from 0.34 to 0.39, and most fields show higher run-to-run variance. This pattern indicates that the Review Agent mainly stabilizes speaker-profile, semantic, affective, and delivery-related decisions through targeted revision rather than through deterministic schema checks.

Removing prior evidence lowers the macro average to 80.07\%, with the largest category-level drops in Acoustic Scene, Affective and Paralinguistic Cues, and Prosody and Delivery: 2.03, 1.71, and 1.51 points. These changes correspond to decreases on background sound and environment in Acoustic Scene, emotion in Affective and Paralinguistic Cues, and speed and pitch in Prosody and Delivery. Caption is slightly higher without priors, but most acoustically grounded fields decline. The result suggests that heterogeneous prior evidence is less important for semantic wording than for acoustic-scene, affective, and prosodic consistency, where it informs both contract construction and multimodal labeling.

Removing recording-level and cross-segment context produces the largest overall decrease, reducing the macro average to 79.79\%. The category drops are concentrated in Semantic Understanding and Speaker Profile, which fall by 4.30 and 2.41 points, while Affective and Paralinguistic Cues drops by 1.20 points. The most affected fields are caption, contextual inference, perceived age, accent, and paralinguistic cues, with decreases of 4.77, 3.83, 3.37, 2.94, and 2.12 points. In contrast, Voice Quality and Articulation and Acoustic Scene change little, and Prosody and Delivery decreases by only 0.76 points. These results show that the Planning Agent's contextual observation is most useful for semantic interpretation, speaker-dependent attributes, and cross-segment paralinguistic consistency, whereas locally audible voice-quality and scene fields rely less on long-context evidence.

\FloatBarrier

\section{Conclusion}
This paper introduced SpeechAnnotator, an open-source context-aware multi-agent framework for segment-aligned multidimensional speech annotation, together with SA-Bench and SA-Eval. Three specialist agents collaborate through a shared segment-level state: the Planning Agent organizes field-relevant evidence and context, the Labeling Agent performs contract-guided multimodal annotation, and the Review Agent selectively revises semantically inconsistent fields through a bounded review loop. Supporting modules provide timeline construction, transcript refinement, and heterogeneous prior evidence, while deterministic output processing produces schema-aligned annotations. On SA-Bench, SpeechAnnotator achieves the lowest character-level timeline error among evaluated systems, with 12.42\% CER and 33.69\% \textit{tcp}CER, and reaches an attribute macro average of 81.27\%, ranking first on seven of the 15 evaluated attributes and second overall among the five attribute-evaluation systems. The ablation results further show that context-aware planning, prior evidence, and the review loop contribute 1.48, 1.20, and 0.85 points, respectively, to the average attribute score. Future work will explore stronger field-specific evidence modeling, adaptive context selection, and more effective review strategies to further improve annotation quality and generalization across languages and domains.
\bibliographystyle{splncs04}
\bibliography{main}

\end{document}